\documentstyle[12pt,preprint,aps]{revtex}

\begin{document}
\preprint{\vbox{\hbox{April 1996}}}
\vskip1cm
\title{Heavy Quark Solitons in the
Nambu--Jona-Lasinio Model}
\vskip1cm
\author{L.\ Gamberg, H.\ Weigel, U.\ Z\"uckert,
and H.\ Reinhardt}
\address{Institute for Theoretical Physics,
T\"ubingen University \\
Auf der Morgenstelle 14,
D-72076 T\"ubingen, Germany}
\maketitle
\vskip1cm

\begin{abstract}
The Nambu--Jona-Lasinio model (NJL) is extended to incorporate 
heavy quark spin-symmetry.  In this model baryons containing
one heavy quark are analyzed as bound-states of light baryons, 
represented as chiral solitons, and mesons containing one heavy quark.  
From related studies in Skyrme type models, the ground-state heavy
baryon is known to arise for the heavy meson in a P--wave 
configuration. In the limit of an infinitely large quark mass 
the heavy meson wave-function is sharply peaked at the center of 
the chiral soliton. Therefore the bound state equation reduces to 
an eigenvalue problem for the coefficients of the operators 
contained in the most general P-wave {\it ansatz} for the heavy 
meson. Within the NJL model a novel feature arises from the 
coupling of the heavy meson to the various light quark states.
In this respect conceptual differences to Skyrme model calculations 
are discovered: The strongest bound state is given by a heavy 
meson configuration which is completely decoupled from the 
grand spin zero channel of the light quarks.
\end{abstract}

\vskip1cm

\leftline{PACS numbers: 12.39.Hg, 12.39.Fe, 12.39.Ki.}

\newpage

\normalsize\baselineskip=20pt
\section{Introduction and Motivation}
\medskip

Recently, there has been considerable attention given to properties 
of hadrons containing a single heavy quark with mass, $M_Q$, much 
larger than the typical scale of strong interactions, $\Lambda_{QCD}$.  
Henceforth these particles will be referred to as ``heavy hadrons".
These ``heavy hadrons" are frequently described within the heavy quark 
effective theory (HQET) \cite{ei81,neu94}, which represents a 
$1/M_{Q}$ expansion of the heavy quark content of QCD. The heavy 
quark effective Lagrangian is set up from the heavy quark 
transformation 
\begin{eqnarray}
Q_{v}^{(l,s)}(x)=\frac{1\pm {v \hskip -0.5em /}}{2}\: 
{\rm{exp}}\:(iM_{Q}v\cdot x)Q(x),
\label{hqt}
\end{eqnarray}
which disentangles the large $(l)$ and small $(s)$ components of the 
heavy quark spinor, $Q(x)$. In eq (\ref{hqt}), $v_{\mu}$ refers to the 
velocity of the heavy quark spinor which, {\it e.g.}~in its rest 
frame takes the form, $v_{\mu}=(1,{\bf 0})$. The Dirac operator for the 
large component is expanded as $v\cdot D+{\cal O}(1/M_{Q})$. Since the 
leading order term commutes with the quark spin operator, heavy mesons 
differing only by their spins, as {\it e.g.} pseudoscalar and vector 
mesons, are degenerate in the heavy quark limit, $M_Q\rightarrow\infty$.
The empirical spectrum of heavy mesons \cite{PDG94}, 
$m_{D}=1.87{\rm GeV}\approx m_{D^{*}}=2.01{\rm GeV}$
and $m_{B}=5.28{\rm GeV}\approx m_{B^{*}}=5.32{\rm GeV}$
represents the most convincing evidence of the heavy quark
symmetry\footnote{See \cite{neu94} for further manifestations
of the heavy quark symmetry.}. This degeneracy is in 
contrast to the spectrum of mesons containing only light quarks. Here
the pseudoscalars are much lighter than the vector mesons.
This is understood by interpreting the light pseudoscalar mesons as
the would--be Goldstone bosons of spontaneous chiral symmetry
breaking. 

The incorporation of the heavy quark symmetry into effective meson 
theories is accomplished by introducing the heavy meson field, 
${\cal H}_{lh}^{a}$, in the heavy quark rest frame \cite{bj90}. Here, 
$l$ and $h$ denote the light and heavy quark spins respectively, while 
$a=u,d$ refers to the isospin of the light quark. This field contains 
pseudoscalar $(P^a)$ and vector $(V^a_\mu)$ mesons on equal footing 
\cite{bj90},
\begin{eqnarray}
{\cal H}_{lh}^{a}=iP^{a}{\delta_{lh}}+{\sigma^{i}_{lh}}V^{a}_{i}.
\label{hwf}
\end{eqnarray}
According to the heavy quark symmetry the heavy spin index ($h$) 
decouples completely in effective theories\footnote{Note that in the 
rest frame of the heavy quark the time component of the vector field 
is suppressed in the heavy quark limit by virtue of the 
Proca--type equation of motion.}. The interactions of 
${\cal H}^a_{lh}$ with the light mesons proceeds via the light 
quark indices $(l,a)$ and is governed by chiral symmetry. Such 
effective models have been employed to not only study the properties 
of heavy mesons, (spectrum, form factors, weak decays) \cite{ei81,neu94}
but also the spectrum of heavy baryons 
\cite{je92,rho93,gu93,sch95a,sch95,min95}. In this context the soliton 
picture for baryons is commonly adopted. Within this approach the 
baryon number is carried by a solitonic configuration of 
the light meson fields. Heavy baryons emerge when a heavy meson is 
bound in the background of this soliton. This picture relies on the 
bound state approach to strange baryons initiated by Callan and 
Klebanov \cite{ca85,bl88}. The main difference here is that for the 
heavy meson both the pseudoscalar and vector meson degrees of 
freedom have to be taken into account as a consequence of the 
above mentioned degeneracy. A continuous transition from 
the hyperons to the heavy baryons containing $c$-- or $b$--quarks 
instead of an $s$--quark suggests a P--wave configuration for the 
bound meson. 

In the heavy quark limit the heavy meson wave--function, 
${{\cal H}_{lh}^{a}}$, is 
confined to the center $(r=0)$ of the soliton. For a quantitative 
determination of this effect it is necessary to take into account the 
interaction amongst the heavy quark degrees of freedom before performing 
the transformation (\ref{hqt}) \cite{sch95}. However, to leading order 
in the $1/M_{Q}$ expansion it is consistent to investigate the 
interaction of a strongly peaked heavy meson wave--function and the 
light quark degrees of freedom while henceforth discarding that part 
of the action which only describes heavy quark degrees of freedom.
Up to ${\cal O}(1/M_Q)$ corrections the binding energy of the heavy 
baryon is thus determined by the properties of the soliton at 
$r\approx 0$ \cite{je92,rho93,gu93}. In that case the bound state 
equation reduces to an eigenvalue problem for the coefficients of 
the operators in the general P--wave {\it ansatz} for 
${{\cal H}_{lh}^{a}}$.

Here we study such bound systems within the Nambu--Jona-Lasinio 
(NJL)  model \cite{nam61} extended to incorporate heavy quark symmetry 
\cite{ebe95}\footnote{Similar studies have been carried out in 
ref \cite{no93}.}. In contrast to Skyrme type models the soliton in the 
NJL model contains components of non-vanishing grand spin. The grand 
spin is the vector sum of total angular momentum and isospin, 
$\mbox{\boldmath $G$}=\mbox{\boldmath $J$}+\mbox{\boldmath $\tau $}/2$.
The appearance of additional modes is due to the quark sub-structure of 
the soliton and leads to a novel coupling scheme between the heavy meson 
and the soliton as will become clear in the course of this paper. In 
particular, the largest binding energy is obtained for a heavy meson 
configuration which completely decouples from the grand spin zero 
channel of the light quarks.

In section~2 we review the extension of the NJL model to include 
heavy quarks. In section~3 the bound state wave--function of the heavy 
meson in the background of the soliton in the NJL model is constructed. 
The numerical results are presented in section~4 and concluding remarks 
are contained in section~5. Some technical details are relegated to 
an appendix.

\bigskip
\section{Heavy Quarks within the NJL Model}
\medskip

In NJL-type quark models the interaction Lagrangian is given by a four 
quark contact term.  The heavy quark symmetry is incorporated by only 
taking into account the large component $Q_{v}\:{\equiv}Q_{v}^{(l)}$ 
which is defined in eq. (\ref{hqt}). Furthermore, that part of the 
Lagrangian, which only contains heavy quarks, $Q$, is discarded. We 
refer to ref \cite{ebe95} for the explicit form of the 
four quark interaction between the heavy and light quarks. Following 
the path integral bosonization procedure \cite{ebe86}, 
a composite field, $H$, is introduced such that the Lagrangian is 
bilinear in the quark fields,
\begin{eqnarray}
{\cal L}^{hl}=
{\bar{Q}_{v}}
(iv\!\cdot \!{\partial}){Q_{v}}
-{\bar{Q}}_{v}H{\tilde q}
-{\bar{\tilde q}}\:{\bar{H}}Q_{v}+ 
\frac{1}{2 {G_{3}}}\:{\rm{tr}}({\bar{H}}H),
\label{lhl}
\end{eqnarray}
where $\tilde q$ denotes the Dirac spinor of the light quarks in the 
chirally rotated representation $\tilde q=\Omega q$ \cite{re89a}. 
Denoting by $\xi$ the root of the chiral field, $U=\xi\xi$ in 
unitary gauge, we have
$\Omega=\xi+\xi^{\dag}+\left(\xi-\xi^{\dag}\right)\gamma_5$. 
The second and third terms in eq. (\ref{lhl}) represent the interaction
between the heavy and light quarks which is mediated by the heavy
meson field, $H=(H^u,H^d)$ \cite{bj90}
\begin{eqnarray}
H^a={\frac{1}{2}}(1+{v \hskip -0.5em /})
(i{\gamma}_{5}P^{a}+{V \hskip -0.7em / \ }^{a}),
\quad {\bar{H}^a}=\gamma_{0}{H^{\dag a}}\gamma_{0}\quad
{\stackrel{{\scriptstyle v_\mu\to(1,{\bf 0})}}
{\textstyle\longrightarrow}}
\pmatrix{0 & 0 \cr {\cal H}^a & 0 \cr},\quad a=u,d.
\label{hmfield}
\end{eqnarray}
Finally the last term in eq. (\ref{lhl}) is purely mesonic.  The
coupling constant $G_{3}$ stems from the four quark interactions
between the heavy and light quark fields.  In eq. (\ref{lhl}) only 
the leading terms of the  $1/M_{Q}$ expansion have been maintained. 
Therefore, the kernel ($iv\!\cdot \!{\partial}$) for $Q_{v}$ is 
that of a massless non-relativistic particle.
Integrating out the fermion fields yields
\begin{eqnarray}
A&=&A_{\cal F}+A^l_m+A^h_m, \nonumber \\
A_{\cal F}&=& {\rm Tr_{\Lambda}} \: \log \;\bf D, \nonumber\\
A^h_m&=&\frac{1}{2{\rm{G_{3}}}}
\int d^{4}x\;{\rm tr}\;\left({\bar H}H\right).
\label{act}
\end{eqnarray}
The functional trace stems from the path--integral over all quark 
fields and involves the inverse propagator,
\begin{eqnarray}
{\bf D}=
\left(
\begin{array}{cc}
i{{D \hskip -0.7em /}\ }^\prime_l & -{\bar{H}} \cr
-H & iv\cdot  \partial \end{array} \right) , \qquad
i{{D \hskip -0.7em /}\ }^\prime_l=
\Omega i{{D \hskip -0.7em /}\ }_l\ \Omega^{\dag}=
\Omega\Big(i\partial \hskip -0.5em / - 
m(U)^{\gamma_5}\Big)\Omega^{\dag}\ ,
\label{mtx}
\end{eqnarray}
which acts on the light-heavy spinor 
$\pmatrix{\tilde q\cr Q_{v}\cr}$. Here $m$ denotes 
the light quark constituent mass. As a consequence of the chiral 
rotation and the omission of the light (axial) vector mesons, 
the light meson fields only appear in $A^l_m$. Its explicit 
form may {\it e.g.} be found in ref. \cite{re89a}. Finally,
${{D \hskip -0.7em /}\ }_l$ refers to the Dirac operator for the light 
quarks, $u,d$ and describes the interaction between these quarks 
and soliton constructed from the light mesons which in turn are 
quark composites. The explicit form of ${{D \hskip -0.7em /}\ }_l$ 
will be presented in section~3.

In eq. (\ref{act}) we have already indicated the need for 
regularization. For the ongoing discussion we will employ Schwinger's 
proper-time regularization \cite{sch51}.  In the NJL model this 
implies a continuation to Euclidean space $(\tau=x_{4}=ix_{0})$. 
This defines the Euclidean Dirac operator ${\bf D}_{E}$ and the 
Euclidean action $A_{\cal F}^{(E)}={\rm Tr}_{\Lambda} \log {\bf D}_{E}$.
Subsequently, its real part is replaced by a parameter integral
\begin{eqnarray}
A_{\cal F}^{R}=\frac{1}{2}{\rm Tr}_{\Lambda}
\log {\bf D}_{E}{\bf D}_{E}^{\dag}&=&-\frac{1}{2} 
\int_{1/\Lambda^2}^\infty\frac{ds}{s}\ {\rm Tr \ exp}
\left(-s{\bf D}_E{{\bf D}_E \hspace{0.04cm}^{\dag}}\right),
\label{areg}
\end{eqnarray}
with the high momentum contribution chopped off. The imaginary part, 
\begin{eqnarray}
A_{\cal F}^{I}&=&\frac{1}{2}{\rm Tr} 
\log {\bf D}_{E}({\bf D}_{E}^{\dag})^{-1}
\label{ireg}
\end{eqnarray}
is finite and remains unchanged. After computing the functional trace 
the action is continued back to Minkowski space. To adjust the 
model parameters we demand the physical pion mass and decay constant, 
$m_\pi=135{\rm MeV}$ and $f_\pi=93{\rm MeV}$, respectively. Then all 
but one parameter of the light quark sector are determined 
\cite{ebe86,alk96}. For transparency this is commonly chosen to be 
the constituent quark mass, $m$. Subsequently, the heavy quark coupling 
constant, $G_3$, is fitted to the $B$-meson decay constant 
${f}_B \approx 180 {\rm  MeV}$ as estimated from lattice QCD 
\cite{al91} and QCD sum rules \cite{co91}. Since the present treatment 
closely follows the computation of ref. \cite{ebe95} we dispense with 
further details although one remark is in order. As mentioned above, 
the light (axial) vector mesons are omitted in the present study. As 
a consequence of the missing $\pi - a_1$ mixing the relation between 
$f_\pi$ and $\Lambda$ is modified whence reducing the numerical value 
of $\Lambda$. This also effects the prediction for $G_3$. 
We display the resulting values in table~\ref{tab_mes}. For the system 
without the light (axial) vector mesons we have re-evaluated the 
binding energy, $E_M=m-\triangle M$, of the heavy meson by determining 
the root, $\triangle M$ of the Bethe-Salpeter equation for the heavy 
meson field $H$. For simplicity we expand this Bethe-Salpeter 
equation in powers of $v\cdot p$, with 
$p_\mu=P_{\mu}-{M_{Q}}{v_{\mu}}$ being the residual four-momentum of 
$H$. The coefficients in this expansion depend on the proper-time 
cut-off, $\Lambda$. It turns out that the expansion in $v\cdot p$ 
converges less rapidly than in the model with light (axial) vector 
mesons. An expansion up to at least cubic order proved necessary to 
gain reliable results. In table \ref{tab_mes} we present the predicted 
root, $\triangle M$, and the binding energy, $E_M$. These results 
were obtained from an expansion up to fifth order in $v\cdot p$. 
The analytic form of this expansion is provide in the appendix 
{\it cf.} eqs (\ref{app2}) and (\ref{app3}). 
Apparently the binding energy increases with the constituent quark mass. 
It should be stressed that these results correspond to the leading order 
of the $1/M_{Q}$ expansion.

Eq. (\ref{mtx}) shows that only the residual momentum, $p_{\mu}$, is
affected by the regularization as a consequence of the heavy quark
transformation (\ref{hqt}). This is physically meaningful and crucial 
for a consistent interpretation of the model. Indeed, the residual
momentum of the heavy quark in a hadron containing a single heavy
quark arises entirely from its interaction with the light degrees 
of freedom (in this case light quarks) and is thus cut off at the same scale
$\Lambda$.   Note that the cut--off $\Lambda$ is 
significantly smaller than the c-- or b-- quark masses 
({\it cf.} table \ref{tab_mes}).

\bigskip
\section{Baryons with a Heavy Quark}
\medskip

Before describing heavy baryons as bound systems it is 
appropriate to briefly explain the emergence of the soliton. When 
integrating out the light quarks an energy functional for the light 
meson fields can be extracted \cite{re89,alk96} by first defining the 
one--particle Dirac Hamiltonian, $h$, from 
$i\beta{{D \hskip -0.7em /}\ }_l=i\partial_t-h$ , 
{\it cf.} eq (\ref{mtx}). 
Adopting the hedgehog {\it ansatz} for the chiral field 
$U={\rm exp}\left(i\mbox{\boldmath $\tau$}\cdot
\hat{\mbox{\boldmath $r$}}\Theta(r)\right)$ this Hamiltonian reads
$h=\mbox{\boldmath $\alpha$} \cdot \mbox{\boldmath $p$}
+m\beta\left({\rm cos}\Theta+i\gamma_5\mbox{\boldmath $\tau$}\cdot
\hat{\mbox{\boldmath $r$}}{\rm sin}\Theta\right)$. The total energy 
functional is expressed in terms of the eigenvalues, $\epsilon_\mu$,
of $h$. Formally it is composed, respectively, of valence, 
vacuum, and meson contributions 
\begin{eqnarray}
E_{\rm tot}[\Theta]=N_C\eta_{\rm val}|\epsilon_{\rm val}| 
+\frac{N_C}{2}\sum_\mu \int_{1/\Lambda^2}^\infty 
\frac{ds}{\sqrt{4\pi s^3}}\ {\rm e}^{-s\epsilon_\mu^2} 
+m_\pi^2f_\pi^2\int d^3r \left(1-\cos\Theta(r)\right).
\label{eng}
\end{eqnarray}
Here $N_C=3$ is the number of color degrees of freedom. Furthermore, 
$\eta_{\rm val}$ denotes the occupation number of the valence quark 
level. This level is defined to be the one with the smallest module 
$|\epsilon_\mu|$. Its occupation number is adjusted to describe 
a configuration possessing unit baryon number, {\it i.e.} 
$\eta_{\rm val}=1+(1/2)\sum_\mu{\rm sgn}(\epsilon_\mu)$. Finally 
the chiral angle, $\Theta(r)$ is self--consistently determined by 
minimizing $E_{\rm tot}$ \cite{re88,alk96}.

In order to compute the binding energy of the heavy meson in the 
soliton background the bosonized action (\ref{act}) is first expanded 
up to the quadratic order in the meson field,
$H(\mbox{\boldmath $r$},t)=\int (d\omega/2\pi)\ 
\tilde H(\mbox{\boldmath $r$},\omega)\ {\rm exp}(-i\omega t)$
\begin{eqnarray}
A_{hl}^{(2)}=\hspace{-0.1cm}\int\frac{d\omega}{2\pi}
\Bigg\{\frac{2\pi}{N_CG_3}{\rm tr}\left(
{\bar{\tilde H}}\tilde H\right)
-\eta_{\rm val}\frac{\langle {\rm val}|\Omega^{\dag}\beta
{\bar{\tilde H}}\tilde H\Omega|{\rm val}\rangle}
{\omega+\epsilon_{\rm val}}
-\sum_\mu\langle \mu|\Omega^{\dag}\beta{\bar{\tilde H}}\tilde H
\Omega|\mu\rangle R_\Lambda(\omega,\epsilon_\mu)\Bigg\},
\label{actexp}
\end{eqnarray}
where the arguments $(\mbox{\boldmath $r$},\omega)$ of the heavy 
fields have been omitted. In analogy to the static energy functional 
(\ref{eng}) this part of the action is composed of a purely mesonic 
piece as well as valence and vacuum contributions. The regularization 
function, $R_\Lambda$ is obtained as an expansion in the frequency 
$\omega$
\begin{eqnarray}
R_\Lambda(\omega,\epsilon)&=&\frac{1}{2\epsilon}
\left(1-{\rm sgn}(\epsilon)\ {\rm erfc}
\left(\left|\frac{\epsilon}{\Lambda}\right|\right)\right)
-\frac{\omega}{2\epsilon^2}\Big(1-{\rm sgn}(\epsilon)\Big)
\nonumber \\ && \hspace{1cm}
+\frac{\omega^2}{2\epsilon^3}\left\{
\left(1-{\rm sgn}(\epsilon)\ {\rm erfc}
\left(\left|\frac{\epsilon}{\Lambda}\right|\right)\right)
-\frac{2\Lambda}{\sqrt{\pi}\epsilon}
\left(1-{\rm e}^{-\epsilon^2/\Lambda^2}\right)\right\}
+\ldots\ .
\label{rlexp}
\end{eqnarray}
This expansion is equivalent to the one in terms of $v\cdot p$ for 
the meson sector. For completeness we list $R_\Lambda$ up to fifth 
in the appendix, {\it cf.} eq (\ref{app5}). It is not surprising that 
this result for $A^{(2)}$ 
can equivalently be obtained from the expression found for the 
fluctuations of kaon fields in the background of the NJL model soliton 
\cite{we93} once the eigenenergies for the strange quark levels are 
set to zero. We note that in the unregularized formulation only the 
negative energy states contribute to the vacuum part of $A^{(2)}$ 
because
\begin{eqnarray}
\lim_{\Lambda\to\infty}R_\Lambda(\omega,\epsilon)=
\frac{1}{2}
\frac{1-{\rm sgn}(\epsilon)}{\omega+\epsilon} .
\label{runreg}
\end{eqnarray}
In the many body interpretation of the functional integral, 
$A_{\cal F}$, one would phrase this result such that in the limit
$\Lambda\to\infty$ only the occupied quark orbits couple to the heavy 
meson. The unregularized expression (\ref{runreg}) may independently 
be derived when considering 
\begin{eqnarray}
A_{hl}^{(2)}({\rm unreg.})={\rm Tr}\left\{
\left(i{{D \hskip -0.7em /}\ }_l\right)^{-1}{\bar H}G_vH\right\}
\label{a2unreg}
\end{eqnarray}
in Minkowski space. Here 
\begin{eqnarray}
G_v(t,t^\prime)=
\langle t|\left(i\partial_t\right)^{-1}|t^\prime \rangle=
\lim_{\eta\to0}\int\frac{dk}{2\pi}\
\frac{{\rm e}^{-ik(t-t^\prime)}}{k+i\eta}
\label{gvmat}
\end{eqnarray}
denotes the non--relativistic propagator in the heavy quark rest 
frame. However, the regularization is unavoidable since 
$\sum_\mu R_\infty(\omega,\epsilon_\mu)$ diverges logarithmically. 
Nevertheless, the limit (\ref{runreg}) serves as an independent check
on our computations. From the expression (\ref{runreg}) it also 
becomes apparent that the total action (\ref{actexp}) is continuous as 
the energy eigenvalue associated with the valence quark orbit changes 
its sign.

Next we study the coupling of a grand spin $\frac{1}{2}$, P--wave 
heavy meson to the soliton. Such a configuration can be expressed in 
terms of three radial functions \cite{sch95}
\begin{eqnarray}
{\tilde {\cal H}}^a_{lh}(\mbox{\boldmath $r$},\omega)=
\hat{\mbox{\boldmath $r$}}\cdot
\mbox{\boldmath $\tau$}^{ab}
\left\{u_A(r,\omega)\delta_{lh}\delta^{bc}
+u_B(r,\omega)\mbox{\boldmath $\tau$}^{bc}\cdot
\mbox{\boldmath $\sigma$}_{lh}
+u_C(r,\omega)\hat{\mbox{\boldmath $r$}}\cdot
\mbox{\boldmath $\tau$}^{bc}
\hat{\mbox{\boldmath $r$}}\cdot
\mbox{\boldmath $\sigma$}_{lh}
\right\}\chi^{c} .
\label{hpwave}
\end{eqnarray}
Here $\chi^a$ denotes a (constant) isospinor. One might have 
expected the appearance of three different isospinors. However, such 
an {\it ansatz} is waived because it is not an eigenstate of the 
grand spin projection operator. Skyrme model studies 
\cite{sch95a} also indicate that the light grand spin, which is 
defined as the grand spin of the operator multiplying $\chi$, has to 
vanish for the meson configuration with the largest binding energy.
In the heavy quark limit the binding energy of this heavy meson is 
determined by the potential it experiences at the center of the 
soliton, $r=0$. In order to extract this potential it is convenient 
(and consistent) to adopt radial functions in eq (\ref{hpwave}) which 
are non--vanishing only at $r=0$. 
Consequently, the heavy meson field only couples to quark levels, 
which have non--vanishing radial functions at $r=0$. As 
can be seen from the decomposition (\ref{hmfield}) and the coupling 
scheme exhibited in the Lagrangian (\ref{lhl}), these non--vanishing 
radial functions must occur in the lower components of the light quark 
fields. At first sight this appears to be surprising because the lower 
component of the valence quark eigenfunction happens to vanish at $r=0$. 
However, one has to be aware of the fact that the heavy meson couples to 
the light quark states via the chiral rotation, which is off--diagonal 
at the origin in the presence of the soliton since $\Theta(r=0)=-\pi$,
\begin{eqnarray}
\Omega(\mbox{\boldmath $r$}=0)=\pmatrix
{0 & i\mbox{\boldmath $\tau$}\cdot \hat{\mbox{\boldmath $r$}}\cr
i\mbox{\boldmath $\tau$}\cdot \hat{\mbox{\boldmath $r$}} & 0 \cr}\ ,
\label{omegar0}
\end{eqnarray}
allowing the valence quark to couple to the heavy meson. In addition 
to the channel with grand spin zero, which includes the valence quark 
orbit, only the channel carrying grand spin one contains states, which 
do not vanish at $r=0$. 

Carrying out the angular integrals as well as the flavor and Dirac 
traces finally yields the action as a function of the values which 
the meson wave--functions in eq (\ref{hpwave}) assume at $r=0$. 
Denoting these by $A(\omega),B(\omega)$ and $C(\omega)$,
respectively, we find
\begin{eqnarray}
A_{lh}^{(2)}=-\frac{1}{2}\int\frac{d\omega}{2\pi}
\left(\chi^{\dag}\chi\right) 
\Big(A^*(\omega),B^*(\omega),C^*(\omega)\Big)\ 
{\cal G}(\omega)
\pmatrix{A(\omega)\cr B(\omega)\cr C(\omega)} .
\label{a2abc1}
\end{eqnarray}
The inverse propagator is a $3\times3$ matrix
\begin{eqnarray}
{\cal G}(\omega)=\frac{8\pi}{N_CG_3}
\pmatrix{1&0&0\cr 0&3&1\cr 0&1&1\cr}
+f_0(\omega)\pmatrix{1&-3&-1\cr -3&9&3\cr -1&3&1\cr}
+f_1(\omega)\pmatrix{3&3&1\cr 3&3&1\cr 1&1&3\cr}.
\label{a2abc2}
\end{eqnarray}
Denoting by $g_\mu(r)$ the radial parts of the upper components of 
the light quark wave--functions, which are eigenstates of $h$, the
spectral functions are expressed as\footnote{The single quark 
wave--functions, $g_\mu$, carry the dimension $({\rm energy})^{3/2}$.}
\begin{eqnarray}
f_0(\omega)&=&\eta_{\rm val}
\frac{\left|g_{\rm val}(0)\right|^2}{\omega+\epsilon_{\rm val}}
+\sum_{\mu,G=0}\left|g_\mu(0)\right|^2 
R_\Lambda\left(\omega,\epsilon_\mu\right),
\nonumber \\
f_1(\omega)&=&
\sum_{\mu,G=1}\left|g_\mu(0)\right|^2 
R_\Lambda\left(\omega,\epsilon_\mu\right).
\label{spectfct}
\end{eqnarray}
The sums refer to distinct quark grand spin ($G$) channels.

The equation of motion for the heavy meson reduces to a 
homogeneous matrix equation for $A,B$ and $C$. Its solution requires 
adjusting $\omega$ to $\omega_0$ such that ${\cal G}(\omega_0)$ 
possesses vanishing eigenvalues. It is easy to show that there exist 
two distinct solutions to this problem
\begin{eqnarray}
&(1):& \frac{2\pi}{N_CG_3}+f_0(\omega_0)=0,\ 
{\rm with}\ (A,B,C)\propto (1,-1,0) \ ,
\label{seta} \\
&(2):& \frac{2\pi}{N_CG_3}+f_1(\omega_0)=0,\
{\rm with}\ (A,B,C)\propto (3+\zeta,1,\zeta) \ .
\label{setb}
\end{eqnarray}
The vectors of set (2) obviously span a two dimensional vector space 
causing the associated eigenstates to be degenerate. The set (1) 
decouples from the $G=1$ states while the set (2) is annihilated by 
the $G=0$ states. In particular, the set (1) corresponds to the 
bound state wave--function found in the heavy quark limit of properly 
extended Skyrme type models \cite{sch95a}
\begin{eqnarray}
{\cal H}^a_{lh}({\rm Skyrme})=u(r)\left(\hat{\mbox{\boldmath $r$}}
\cdot \mbox{\boldmath $\tau$}\right)_{al}
\left(\tau^2_{hh^\prime}\chi_{h^\prime}\right)
=\frac{u(r)}{2}\left\{\hat{\mbox{\boldmath $r$}}\cdot
\mbox{\boldmath $\tau$}\left(\delta_{lh}
-\mbox{\boldmath $\tau$}\cdot \mbox{\boldmath $\sigma$}_{lh}\right)
\chi\right\}_a .
\label{hskyrme}
\end{eqnarray}
The interpretation of this situation is obvious: If the background 
soliton field contains only $G=0$ states, as it is the case 
in Skyrme type models, the only solution will be given by the set (1). 
However, if also $G=1$ states are present an alternative 
solution will exist. The soliton dynamics, from which the spectral 
functions $f_0$ and $f_1$ are computed, then determines which of these 
two modes, (1) or (2), leads to a larger binding and therefore has to be
interpreted as the lightest baryon containing a heavy quark. 

\bigskip
\section{Numerical Results for the Heavy Baryon}
\medskip

In analogy with the meson sector we have expanded the regularization 
function, $R_\Lambda(\omega,\epsilon)$, (\ref{rlexp}) up to fifth 
order in the frequency $\omega$. In figure 1 the resultant 
frequency dependence of the spectral functions (\ref{seta}) and 
(\ref{setb}) is shown for the special case $m=400{\rm MeV}$. 
As a result of the heavy quark transformation (\ref{hqt}) the threshold
for the (unphysical) decay of the heavy meson into a quark--antiquark 
pair is given by the energy of the lowest accessible light quark state. 
Therefore a singularity appears at $\omega\approx-\epsilon_{\rm val}$ 
for the mode coupling to the $G=0$ channel, while the radius of 
convergence is much larger in the case of the $G=1$ one channel. The 
latter mode apparently leads to a smaller eigenfrequency, $\omega_0$.

The total energy of the system consisting of the soliton and the bound 
meson is given by $M_Q+E_{\rm tot}+\omega_0$. As the total energy of 
the heavy meson in the trivial background is $M_Q+\triangle M$ the 
binding energy of the heavy baryon becomes $E_B=\triangle M-\omega_0$. 
This binding energy is measured with regard to a decay into a soliton 
and a heavy meson. When increasing the heavy quark coupling constant, 
$G_3$, the curves shown in figure 1 get shifted downwards 
by a constant amount. Hence the roots decrease and the binding energies
of the baryon states containing a heavy quark become larger when the 
coupling gains strength. Obviously, a heavy baryon constructed from 
an eigenmode of type (2) is more strongly bound than a type (1) heavy 
meson. In table \ref{tab_bar} the predicted roots, $\omega_0$, as well 
as the associated binding energies, $E_B$, of the heavy baryon are 
presented.
For the modes of type (1) the eigenfrequency acquires a maximum
for $m\approx 500{\rm MeV}$. Since $\triangle M$ increases with 
the constituent quark mass, $m$, the binding energy for the associated 
heavy baryon exhibits almost no variation for $m\le500{\rm MeV}$. 
When $m$ is further increased the eigenfrequency decreases leading 
to a larger binding energy. This behavior of $\omega_0$ is due to 
the decrease of $\epsilon_{\rm val}$ when $m$ grows. On the contrary, 
the eigenfrequency corresponding to the eigenvectors of type (2) 
appears to be a quickly, monotonously decreasing function of $m$. 
This causes the binding energy of the corresponding heavy baryon to 
significantly increase with $m$. Since studies in the light quark 
sector of the NJL model favor $400{\rm MeV}\le m\le 450{\rm MeV}$ 
\cite{alk96} the results displayed in table \ref{tab_bar} suggest 
the interpretation that the NJL model predicts the baryon with one 
heavy quark to have a binding energy of about $560{\rm MeV}$ in 
leading order of the $1/M_Q$ expansion. A less strongly bound state 
is obtained possessing about half of that binding energy. As 
discussed above the latter state corresponds to the one which is 
identified as the most strongly bound baryon in Skyrme type models. 
In this context it has to be remarked that a state associated with 
the mode (2) does not appear in Skyrme type models because in those 
models the background soliton is a pure $G=0$ configuration. If the 
NJL model soliton had contained no $G=1$ components the most strongly 
bound baryon would have been of the same structure as in purely 
mesonic soliton models.

As in the bound state approach \cite{ca85} heavy baryon states with 
good spin and isospin are generated by canonically quantizing the 
collective isospin rotation, $A(t)$, of the meson fields. This adds a 
term of the form $C_i{\rm tr}(\chi\chi^{\dag}A^{\dag}(t){\dot A}(t))$
to the action. The index $i=1,2$ refers to the distinct sets 
in eqs (\ref{seta},\ref{setb}). The time derivative of the 
collective rotation defines the angular velocity 
$\mbox{\boldmath $\omega$}$ via
$A^{\dag}(t){\dot A}(t)=(i/2)\mbox{\boldmath $\omega$}
\cdot \mbox{\boldmath $\tau$}$. Empolying (iso)rotational 
invariance these coefficients can hence be computed from the matrix 
elements\footnote{For the case of the bound state approach to 
hyperons these matrix elements have thoroughly been discussed 
by Weigel {\it et al.} in ref \cite{bl88}.}
$\langle \mu |\tau_3|\nu\rangle\langle \nu|
\Omega^{\dag}H^{\dag}\beta H\Omega|\mu\rangle$. As argued above
$H$ only couples to quark states with $G=0,1$. The 
bound state of set (1) has non--vanishing matrix elements only 
for quark levels with $G=0$. In that case the matrix element 
of $\tau_3$ vanishes, hence $C_1=0$, in agreement with the Skyrme 
model results \cite{sch95a}. For the set (2) there is indeed a 
coupling when both quark levels ($|\mu\rangle$ and $|\nu\rangle$)
are from the $G^\pi=1^+$ channel. We find
$C_2={\cal C}_2[(\zeta+2)(\zeta+6)]/[(\zeta+2)^2+2]$, where the 
positive definite denominator is due to the normalization 
associated to the metric induced by the coefficient matrix of $f_1$ 
in eq (\ref{a2abc2}). Although the collective rotation removes the
degeneracy in $\zeta$ we will argue that ${\cal C}_2$ is negligibly 
small without going into the details of the calculation. For quark
states ($|\mu\rangle$ and $|\nu\rangle$) from the $G^\pi=1^+$ channel 
the above 
mentioned matrix elements are similar to those entering the evaluation
of the moment of inertia $\alpha^2$ \cite{re89}. As the collective 
quantization gives rise to a factor $1/\alpha^2$, ${\cal C}_2$ can
be estimated by the contribution of the $G^\pi=1^+$ channel to the 
total moment of inertia. This contribution is very small because 
$\alpha^2$ is dominated by the valence quark level, which resides in 
the $G^\pi=0^+$ channel, {\it e.g.} for $m=400{\rm MeV}$ we find 
a contribution of only 2\%. Hence we conclude that there is (almost)
no hyperfine splitting in the heavy quark limit. For a finite heavy 
quark mass this conclusion should change because $H$ will couple to 
channels with $G\ge2$ as well. 

\bigskip
\section{Conclusions}
\medskip

We have analyzed fluctuations of mesons with a heavy quark in the 
background of the NJL model chiral soliton in the leading order of 
the heavy quark mass expansion. We have succeeded in showing that
at this order the associated Bethe--Salpeter equation possesses the 
bound state solution known from Skyrme model studies. We have 
furthermore seen that this Bethe--Salpeter equation contains an 
additional solution which arises from the coupling of the heavy 
meson field to vacuum quark states. Also, the novel bound state 
appears in degenerate pairs. The actual computation reveals that in 
the NJL model the novel states are more strongly bound than those 
configurations known from the Skyrme model. Although this certainly 
represents an interesting feature in its own, future studies are 
required to find out whether or not this novel state is an artifact 
of applying the well--established $1/M_Q$ expansion to the NJL model. 
In comparison a few aspects of the Skyrme model calculation for finite
$M_Q$ \cite{sch95,min95} are worthwhile to be mentioned. First, the 
restriction to the leading order of the $1/M_Q$ expansion has been 
found to be a somewhat crude approximation to the realistic cases 
causing states, which are predicted to be degenerate at leading 
order, to acquire distinct binding energies when corrections of 
subleading order are taken into account. We assume that a similar 
mechanism will remove the degeneracy of the novel states in the NJL 
model. Second, table (1) of ref \cite{sch95} shows that the 
transition from infinite to realistic values of $M_Q$ is indeed 
continuous. In particular the order of bound states remains unchanged. 
Hence one is inclined to conjecture that the novel states persist for 
finite $M_Q$. Third, not all bound states, which are predicted within 
the large $M_Q$ limit are permitted for finite $M_Q$ because the limits
$M_Q\to\infty$ and $r\to0$ do not necessarily commute. This respresents
the only mechanism which would prohibit the existence of the novel 
states in the case of finite $M_Q$. These issues certainly raise interest 
for generalizing the NJL model studies to finite $M_Q$. Such studies are 
considerably more involved because it is crucial to first perform the 
heavy quark transformation and to subsequently regularize the functional 
trace to have a consistent interpretation of the model.

In practice the binding energies of heavy baryons are measured with 
respect to the decay into a nucleon and the lightest meson containing 
the corresponding heavy quark. The experimentally observed binding
energies are $\sim613\pm50{\rm MeV}$ and $\sim625{\rm MeV}$ for the
$\Lambda_B(5641\pm50)$ and $\Lambda_C(2285)$, respectively \cite{PDG94}.
It should be noted that these baryons are degenerate in the heavy quark
limit. In reasonable agreement with these data our numerical analysis 
yields a binding energy of about $560{\rm MeV}$ for the most strongly 
bound heavy baryon when the only free parameter of the model is adjusted
to reproduce the properties of the light baryons. The less strongly bound 
heavy baryon with a predicted binding energy around $250{\rm MeV}$ may 
eventually be associated with $\Lambda_C(2625)$. This baryon is bound 
by about ($\sim285{\rm MeV}$) against the decay into a nucleon and a 
$D$--meson. We would like to emphasize that the main purpose of the 
present paper was to discuss the novel coupling scheme between the 
heavy meson and the soliton rather than to achieve a precise agreement 
with the experimentally observed binding energies. 

\bigskip
\acknowledgements
One of us (HW) is grateful for illuminating discussions with 
Joseph Schechter.  This work is supported by the Deutsche
Forschungsgemeinschaft (DFG) under contract Re 856/2-2. 
HW acknowledges support by a Habilitanden--scholarship of the DFG. 
UZ is Member of the Graduiertenkolleg ``Struktur und Wechselwirkung 
von Hadronen und Kernen'' of the DFG (contract Mu 705/3).

\vskip1cm
\centerline{\bf APPENDIX}
\vskip.5cm

In this appendix we will briefly provide the expansions we have been 
using for the regularization functions, which enter the Bethe--Salpeter 
equations for the heavy quark meson. 

\vskip.25cm

In case no soliton is present these functions are obtained from the 
quark--loop in the self--energy, $\Pi(v\cdot p)$ ({\it cf.} eq (47) 
of ref \cite{ebe95})
\begin{eqnarray}
{\rm tr}\left[{\bar H}\Pi(v\cdot p)H\right]=
-iN_C\int \frac{d^4k}{(2\pi)^4} 
\frac{{\rm tr}\left[\left(k \hskip -0.5em /
-p\hskip -0.5em /+m\right) {\bar H}H\right]}
{\left((k-p)^2-m^2\right)\left(v\cdot p+i\epsilon\right)}.
\label{app1}
\end{eqnarray}
Again, $v$ denotes the velocity of the heavy quark inside the meson 
while $p$ labels the momentum of the meson. We refer to the literature 
\cite{no93,ebe95} on the treatment of this quantity in Euclidean space 
and regularization of the quark loop. When $v\cdot p$ is smaller than
the quark--antiquark threshold, which in the heavy quark limit is 
idential to the light quark constituent mass, the self--energy has 
the Taylor expansion
\begin{eqnarray}
\Pi(v\cdot p)= \sum_{n=0}^\infty \Pi^{(n)}(0)(v\cdot p)^n \ ,
\label{app2}
\end{eqnarray}
where the superscript refers to the deriavtive with respect to 
$v\cdot p$. In the proper--time regularization scheme of ref 
\cite{ebe95}, which actually defines the model, the first five 
coefficients of the expansion (\ref{app2}) are
\begin{eqnarray}
\Pi^{(0)}(0)&=&
\frac{N_Cm^2}{16\pi^2}\Bigg\{
\Gamma\left(-1,\frac{m^2}{\Lambda^2}\right)
+2\frac{\Lambda\sqrt{\pi}}{m}\ 
{\rm exp}\left(-\frac{m^2}{\Lambda^2}\right)
-2\pi\ {\rm erfc}\left(\frac{m}{\Lambda}\right)\Bigg\}\ ,
\nonumber \\
\Pi^{(1)}(0)&=&
\frac{N_Cm}{8\pi^2}\Bigg\{
\Gamma\left(0,\frac{m^2}{\Lambda^2}\right)
+\frac{\Lambda\sqrt{\pi}}{m}\ 
{\rm exp}\left(-\frac{m^2}{\Lambda^2}\right)
-\pi\ {\rm erfc}\left(\frac{m}{\Lambda}\right)\Bigg\}\ ,
\nonumber \\
\Pi^{(2)}(0)&=&
\frac{N_C}{16\pi^2}\Bigg\{
2\Gamma\left(0,\frac{m^2}{\Lambda^2}\right)
+\pi\ {\rm erfc}\left(\frac{m}{\Lambda}\right)\Bigg\}\ ,
\nonumber \\
\Pi^{(3)}(0)&=&
\frac{N_C}{16m\pi^2}\Bigg\{
\frac{4}{3}{\rm exp}\left(-\frac{m^2}{\Lambda^2}\right)
+\pi {\rm erfc}\left(\frac{m}{\Lambda}\right)\Bigg\}\ ,
\nonumber \\
\Pi^{(4)}(0)&=&
\frac{N_C}{16m^2\pi^2}\Bigg\{
\left(\frac{4}{3}+\frac{\sqrt{\pi}m}{2\Lambda}\right)
{\rm exp}\left(-\frac{m^2}{\Lambda^2}\right)
+\frac{\pi}{4}{\rm erfc}\left(\frac{m}{\Lambda}\right)\Bigg\}\ ,
\nonumber \\
\Pi^{(5)}(0)&=&
\frac{N_C}{16m^3\pi^2}\Bigg\{
\left(\frac{8}{15}\left[1+\frac{m^2}{\Lambda^2}\right]
+\frac{\sqrt{\pi}m}{2\Lambda}\right)
{\rm exp}\left(-\frac{m^2}{\Lambda^2}\right)
+\frac{\pi}{4}{\rm erfc}\left(\frac{m}{\Lambda}\right)\Bigg\}\ .
\label{app3}
\end{eqnarray}
We then obtain the binding energy, $\triangle M$, of the 
heavy meson by determining the root from the trunctated expansion
and including the contribution from the purely mesonic part of the 
action, $A_m^h$, (\ref{act}) 
\begin{eqnarray}
-\frac{1}{2G_3}+\sum_{n=0}^5 \Pi^{(n)}(0)(\triangle M)^n=0.
\label{app4}
\end{eqnarray}
The decay constant, $f_H$, of the meson with the heavy quark is obtained
by coupling external electro--weak sources \cite{ebe95}. This then
relates $f_H$ to the field normalization, $\sqrt{\Pi^{(1)}(0)}$
\begin{eqnarray}
f_H=\frac{1}{G_3 \sqrt{\Pi^{(1)}(0)M_H}}
\label{app5}
\end{eqnarray}
where $M_H$ refers to the mass of the heavy meson. Using the 
B--meson parameters $f_B=180{\rm MeV}$ and $M_B=5.3{\rm GeV}$ yields 
the numerical results for $\triangle M$ and $G_3$ listed in table 
\ref{tab_mes}.

As the procedure leading to the regularization function 
$R_\Lambda(\omega,\epsilon)$ has already been explained in section 3,
it sufficiecs to just list the expansion up to order $\omega^5$
\begin{eqnarray}
R_\Lambda(\omega,\epsilon)&=&\frac{1}{2\epsilon}
\left(1-{\rm sgn}(\epsilon)\ {\rm erfc}
\left(\left|\frac{\epsilon}{\Lambda}\right|\right)\right)
-\frac{\omega}{2\epsilon^2}\Big(1-{\rm sgn}(\epsilon)\Big)
\label{app6}\\  && \hspace{-1.0cm}
+\frac{\omega^2}{2\epsilon^3}\left\{
\left(1-{\rm sgn}(\epsilon)\ {\rm erfc}
\left(\left|\frac{\epsilon}{\Lambda}\right|\right)\right)
-\frac{2\Lambda}{\sqrt{\pi}\epsilon}
\left(1-{\rm e}^{-\epsilon^2/\Lambda^2}\right)\right\} 
-\frac{\omega^3}{2\epsilon^4}\Big(1-{\rm sgn}(\epsilon)\Big)
\nonumber \\ && \hspace{-1.0cm}
+\frac{\omega^4}{2\epsilon^5}\Bigg\{
\left(1-{\rm sgn}(\epsilon)\ {\rm erfc}
\left(\left|\frac{\epsilon}{\Lambda}\right|\right)\right)
+\frac{4\Lambda^3}{\sqrt{\pi}\epsilon^3}
\left(1-{\rm e}^{-\epsilon^2/\Lambda^2}\right)
-\frac{4\Lambda}{\sqrt{\pi}\epsilon}\Bigg\}
-\frac{\omega^5}{2\epsilon^6}\Big(1-{\rm sgn}(\epsilon)\Big)\ .
\nonumber
\end{eqnarray}

\newpage

\begin{table}
\caption{
The proper--time cut--off, $\Lambda$, the heavy quark coupling 
constant, $G_3$, the root of the Bethe--Salpeter equation 
$\triangle M$, and the heavy meson binding energy, $E_M$ as 
functions of the constituent quark mass $m$.}
\vspace{1cm}
\begin{tabular}{lddddd} 
$m$ (MeV)
& 350    & 400  & 450  & 500  & 600  \\ \hline
$\Lambda$ (MeV)
& 641   & 631   & 633  & 642  & 672  \\ 
$G_3$ $(10^{-5}$MeV$^{-2})$
& 1.53   & 1.68 & 1.82 & 1.94 & 2.17 \\
$\triangle M$ (MeV)
& 327   & 353   & 369  & 401  & 449  \\
$E_M$ (MeV)
& 23   & 47   & 81  &  99  & 151 \\ 
\end{tabular}
\label{tab_mes}
\end{table}

\vskip2cm

\begin{table}
\caption{
The roots, $\omega_0$, of the spectral functions (\protect\ref{seta}) 
and (\protect\ref{setb}) for the heavy meson modes which couple to the 
grand spin zero and one quark channels, respectively. 
$E_B=\triangle M-\omega_0$ represents the associated binding 
energy of the heavy baryon. All data are in MeV.}
\vspace{1cm}
\begin{tabular}{ldddddd}
 \multicolumn{2}{l}{$m$}
& 350    & 400  & 450  & 500  & 600  \\ \hline
Set (1) & $\omega_0$ 
& 94   & 102   & 113  & 114  & 86\\
& $E_B$ & 233 & 251 & 257 & 287 & 363 \\
\hline
Set (2) & $\omega_0$ 
&-128  & -182 & -223 & -261 & -341 \\
& $E_B$  & 455  & 535  & 592 & 662  & 790 \\
\end{tabular}
\label{tab_bar}
\end{table}

\vskip3cm

\centerline{\bf \large Figure Caption}
\vskip1cm
Fig. 1: The spectral functions and their roots
for the eigenvectors defined in eqs (\protect\ref{seta}) and
(\protect\ref{setb}). The constituent quark mass has been taken
to be $m=400{\rm MeV}$.
\end{document}